\documentclass[doublecol]{epl2}

\begin{document}

\title{On the maximum mass of hyperonic neutron stars}

\author{\'E. Massot\inst{1,2} \and J. Margueron\inst{1}, \and G. Chanfray\inst{3}}

\institute{
\inst{1} Institut de Physique Nucl\'eaire, IN2P3-CNRS and Universit\'e Paris-Sud, F-91406 Orsay CEDEX, France \\
\inst{2} D\'epartement de physique, \'Ecole Normale Sup\'erieure, 24 rue Lhomond, 75231 Paris Cedex 05, France,\\
\inst{3} IPN Lyon, Universit\'e de Lyon, Univ. Lyon 1, CNRS/IN2P3, UMR5822, F-69622 Villeurbanne Cedex}

\pacs{97.60.Jd}{}
\pacs{26.60.Kp}{}

\abstract{Chiral Lagrangian and quark-meson coupling models of hyperon matter
are used to estimate the maximum mass of neutron stars. 
Our relativistic calculations include, for the first time, both Hartree and Fock contributions in a consistent manner.
Being related to the underlying quark structure of baryons, these models are
considered to be good candidates for describing the dense core of neutron stars.
Taking account of the known experimental constraints at saturation density, the
equations of state deduced from these relativistic approaches cannot
sustain a neutron star with a mass larger than 1.6-1.66~$M_\odot$.}

\maketitle


Neutron stars are the most compact stellar objects, and within them, particles that are unstable on earth could be stabilized
at densities above 10$^{14}$~g~cm$^{-3}$.
The equation of state (EoS) at these densities is still largely unknown due to the poor experimental knowledge on
hadron dynamics.
However, the recent measurements of the millisecond pulsars PSR J1614-2230
extend the maximum observed mass from
(1.67$\pm$0.02)~M$_\odot$~\cite{ransom2005}
to
(1.97$\pm$0.04)~M$_\odot$~\cite{DPR10}. This imposes new constrains on the
EoS of dense matter~\cite{lattimer2010}.
A major question is whether the EoS including hyperons can be ruled out 
by this new observation.
Relativistic mean-field~\cite{Schafner-Bielich2002,Wu2011} and 
Brueckner-Hartree-Fock~\cite{Baldo2000,Schulze2006,Schulze2011} models have predicted that the effect of hyperons 
in dense matter is to soften the EoS and, therefore, to lower the maximum mass.
A consistent treatment of the Fock term in a relativistic approach is however missing.
In this Letter we analyze the impact of the Fock term in two effective,
relativistic meson-exchange models related to the quark nature of hadrons: the chiral Lagrangian 
model~\cite{MC08} and the quark meson coupling (QMC) model~\cite{RGM08}.

These models are based on a relativistic meson-exchange theory of the Walecka
type, where the mean-field properties of nuclear matter are
calculated in the Hartree--Fock approximation~\cite{BMG87}. 
Considering scalar $s$, $\omega$, $\delta$, $\rho$ and $\pi$ mesons as well as
nucleons and hyperons, the Lagrangian can be written with the usual notations~\cite{RGM08}
\begin{eqnarray}
\mathcal{L} &=& \sum_f\bar{\psi}_f(i\gamma^\mu\partial_\mu-M_f(s))\psi_f -V(s) +\frac12\partial^\mu s
\partial_\mu s \nonumber \\
&-& \sum_f\bar{\psi}_fg_\omega^f\omega^\mu\gamma_\mu\psi_f
+\frac12m_\omega^2\omega_\mu\omega^\mu -\frac14F^{\mu\nu}F_{\mu\nu} \nonumber \\
&-& g_\delta\sum_f\bar{\psi}_f\vec{\delta}\cdot\vec{I}\psi_f
-\frac12m_\delta^2\vec{\delta}\cdot\vec{\delta}
+\frac12\partial^\mu\vec{\delta}\cdot\partial_\mu\vec{\delta}  \nonumber \\
&-&g_\rho\vec{\rho}^\mu\cdot\sum_f\bar{\psi}_f\gamma_\mu\vec{I}\psi_f
-g_\rho\frac{\kappa_\rho}{2M_N}\partial_\nu\vec{\rho}_\mu\cdot\sum_f\bar{\psi}_f\bar{\sigma}^{\mu\nu}\vec{I}\psi_f
\nonumber \\
&+&\frac{1}{2}m_\rho^2\vec{\rho}_\mu\cdot\vec{\rho}^\mu
-\frac14\vec{F'}^{\mu\nu}\cdot\vec{F'}_{\mu\nu}  +\frac{g_A}{2f_\pi}\frac35\partial_\mu\vec{\pi}\cdot
\sum_f\bar{\psi}_f\vec{G}_T^\mu\psi_f \nonumber \\
&+&\frac12m_\pi^2\vec{\pi}\cdot\vec{\pi} -\frac12\partial^\mu\vec{\pi}\cdot\partial_\mu\vec{\pi} ,
\label{eq:lagrangian}
\end{eqnarray}
where the sum over the flavor index $f$ means a summation over the baryons: N, $\Lambda$, $\Sigma$, $\Xi$.
The isospin operator $\vec{I}$ and the Gamow--Teller operator $\vec{G}_T^\mu$ are given in Ref.~\cite{RGM08}.

\begin{table*}[t]
\setlength{\tabcolsep}{.1in}
\renewcommand{\arraystretch}{1.5}
  \caption{Parameters of chiral and QMC models and saturation properties: density ($n_0$), 
  binding energy ($B_0$), incompressibility ($K_0$), symmetry energy ($a_s$), symmetry energy
  slope ($L_0$) and in-medium Dirac mass for nucleons ($M_N^*$). QMC700 is given in Ref.~\cite{RGM08}.}
  \label{tab:parameters}
\begin{center}
\footnotesize\rm
  \begin{tabular}{ccccccccccccc}
    \hline\hline
    Model  &$m_s$ & $g_s$&$g_\omega$&$g_\rho$&$\kappa_\rho$&$C$& $n_0$ & $B_0$ & $K_0$&$a_s$&$L_0$&$M^*_N/M_N$ \\
    & &  &  &  & (fm$^{-3}$) & (MeV) &  (MeV) & (MeV) & (MeV) & (MeV) & (MeV) & \\
    \hline
    MC1-H    &800 &10.0   &5.06   & 4.14  & 0.0 &1.44 & 0.160 & -15.85 & 248 & 30.0 & 79.4& 0.89\\
    MC1-HF  &880 &10.0  &7.09   & 2.65  &4.2&2.00 & 0.160 & -16.01 & 274 & 30.3 & 66.9 &0.81\\
    MC2-HF  &860 &10.0  &7.31   & 2.65  &4.2&1.90 & 0.160 & -16.05 & 276 & 30.6 & 67.9 &0.80 \\
    MC3-HF  &900 &10.0  &6.33   & 2.65  &5.6&1.92 & 0.160 & -15.91 & 274 & 30.2 & 63.1 & 0.83\\
    MC4-HF  &880 &10.0  &6.57   & 2.65  &5.6&1.83 & 0.160 & -15.92 & 274 & 30.5 & 64.1 & 0.82 \\
    \hline
    QMC700 &700 & 11.94 & 10.66 & 4.167 & 0.0 &0.51 & 0.159 & -15.74 & 338&29.9 & 53.3 & 0.74 \\
    QMC-H &700 & 10.52 & 8.39 & 4.167 & 0.0 & 0.40 & 0.160 & -15.83 & 292 & 32.7 & 89.5 & 0.78 \\
    QMC-HF1 &700 & 11.90 & 10.62 & 4.167 & 0.0 & 0.51 & 0.160 & -15.72 & 338 & 30.4 & 54.4 & 0.77 \\
    QMC-HF2 &700 & 11.30 & 9.31 & 2.87 & 0.0 & 0.46 & 0.160 & -15.74 & 360 & 29.8 & 72.2 & 0.67 \\
    QMC-HF3 &700 & 9.78 & 6.84 & 3.75 & 0.0 & 1.00 & 0.160 & -16.00 & 285 & 31.4 & 69.2 & 0.79 \\
    \hline\hline
  \end{tabular}
\end{center}
\end{table*}

In the chiral Lagrangian model~\cite{MC08}, the meson fields are related to the properties of the
QCD condensate.
Scalar $s$ and pseudo-scalar $\pi$ fields are associated with the fluctuations of the chiral 
quark condensate~\cite{CEG01}, related to the non-linear "Mexican hat" potential,
\begin{eqnarray}
V(s)  &=& \frac{m_s^2}{2}s^2+\frac{m_s^2-m_\pi^2}{2f_\pi}s^3
+\frac{m_s^2-m_\pi^2}{8f_\pi^2}s^4,
\label{eq:scalar_pot}
\end{eqnarray}
deduced from chiral spontaneous symmetry breaking.
In this model, the chiral partner $A_1$ of the $\rho$ meson, being too heavy, is ignored 
and the $\omega$ meson is chiral singlet. 
The first chiral models failed to reproduce the saturation properties of nuclear matter~\cite{CEG01}.
A better description for the saturation is obtained by introducing the scalar polarizability of the 
nucleon $\kappa_{NS}$ in $M_f(s)$~\cite{KM75,BT01},
\begin{eqnarray}
M_f(s) &=& M_f+g_s w^f_s s+\frac12\kappa_{NS}\tilde{w}^f_s s^2,
\label{eqn:Mf}
\end{eqnarray}
which reflects the effects of confinement as originally proposed in Ref.~\cite{Gui88}.
The weight factors $w^f_s$ and $\tilde{w}^f_s$ will be discussed later.

A further extension of the model consists in replacing $\kappa_{NS}$ in Eq.~(\ref{eqn:Mf}) by 
$\kappa_{NS}(s) = \kappa_{NS}(1+(2s)/(3f_\pi))$ where $f_\pi$ is 
the decay constant of the pion~\cite{MC08}. 
Another extension is related to the $\rho$ and $\pi$ tensor interactions, where derivative couplings
lead to a contact term that can be partially or totally suppressed by short-range correlations~\cite{BMG87,MC08}.
We consider chiral models at the Hartree approximation (such as MC1-H in table~\ref{tab:parameters}) as well as
in the Hartree--Fock approximation (such as MC$i$-HF in table~\ref{tab:parameters}).
The index $i$ in MC$i$-H(F) models refers to different combinaisons of the
extensions discussed: 
($i$=1), constant $\kappa_{NS}$ in Eq.~(\ref{eqn:Mf}) and totally suppressed contact term; 
($i$=2) $\kappa_{NS}(s)$ in Eq.~(\ref{eqn:Mf}) and totally suppressed contact term; 
($i$=3) constant $\kappa_{NS}$ and partially suppressed contact term; 
($i$=4) $\kappa_{NS}(s)$ and partially suppressed contact term.
More details on these extensions can be found in Ref.~\cite{MC08}.

In the quark-meson coupling model QMC700 the scalar potential~(\ref{eq:scalar_pot}) is quadratic, 
$V(s) =  m_s^2s^2/2$, at variance with the chiral Lagrangian model~(\ref{eq:scalar_pot});
the scalar polarizability of the nucleon $\kappa_{NS}$ is kept constant as in 
Eq.~(\ref{eqn:Mf}); and only $s$, $\omega$ and $\rho$ vector mesons are considered ($\kappa_\rho=0$). 
The other extensions presented here-before are disregarded.
In Ref.~\cite{RGM08}, self-energies are calculated in the Hartree approximation while a non-relativistic 
approximation of the Fock contribution to the energy is included. 
In this paper, we present modified versions of the QMC model in which both Hartree and Fock
contributions are treated consistently.

The parameters of chiral Lagrangian and QMC models are given in table~\ref{tab:parameters}.
In the chiral model, the scalar coupling constant is fixed from the sigma linear
model to be $g_s=M_N/f_\pi=10$.
Since the dimensionless parameter $C = f_\pi^2\kappa_{NS}/(2M_N)$ estimated from lattice simulation 
is around 1.25~\cite{TGL04}, the chiral model fixes $C$ to be consistent with this value
but a flexibility is retained for reproducing saturation properties as explained below.
In the QMC model, the dimensionless parameter $C$ is smaller and deduced from the nucleon mass in the bag model.
The coupling constant $g_\rho$ is adjusted to fit the symmetry energy in the chiral Hartree model MC1-H and in
QMC models while in chiral Hartree--Fock models MC$i$-HF $g_\rho$ is fixed to satisfy the vector-dominance 
model giving $g_\rho=2.65$ and the $\rho$-tensor coupling constant $\kappa_\rho$ is adjusted to the symmetry energy.
The axial coupling constant is fixed to be $g_A=1.25$ and the $\delta$ coupling is fixed to $g_\delta=1$.
The other parameters ($m_s$, $g_\omega$ and $C$) are chosen to reproduce the saturation properties ($n_0$, $B_0$ and $K_0$) 
reported on table~\ref{tab:parameters}.
In QMC model, the couplings $g_s$ and $g_\omega$ are fixed to reproduce $n_0$ and $B_0$; moreover in the parameterization 
QMC-HF3 the parameter $C$ is modified to decrease the value of $K_0$.

The predictions for the slope of the symmetry energy $L_0$ and the nucleon Dirac mass $M_N^*$ given in 
table~\ref{tab:parameters}
for the two models are consistent with experimental values lying in the range 60-90 MeV~\cite{VPP09} for the slope and in the
range 0.7-0.9$M_N$ for the Dirac masses.

Hyperons  are included as in Ref.~\cite{RGM08}: the coupling constants are written as $g_m^f=w_m^f g_m$ 
where $w_m^f$ is the weight factor for the meson $m$ coupled to the baryon $f$.
For the mesons $\omega$, $\rho$, $\pi$ and $\delta$, the meson-hyperon couplings are deduced from the 
meson-nucleon couplings based on $SU(6)$ symmetry imposing
$w_\omega^f=(1+s^f/3)$, where $s^f$ is the 
strangeness of the baryon $f$.
For the isovector $\rho$ or pseudo-vector $\pi$ mesons, this symmetry is accounted for by the isospin and 
Gamow--Teller operators. 
Only the coupling constants of the scalar $s$ meson are allowed to vary around the values imposed by
$SU(6)$ symmetry.
In QMC700, the weights $w_s^f$ and $\tilde{w}_s^f$ are adjusted in the bag model and scale 
with the bag radius~\cite{RGM08}.
In the chiral model and in the extensions of QMC in table~\ref{tab:parameters}, 
they are adjusted to the non-relativistic potentials of hyperons at saturation, 
\begin{equation}
V_{NR}(f) =
\Sigma_S^f+\frac{E^f}{M^f(s)}\Sigma_0^f+\frac{\Sigma_S^{f2}-\Sigma_0^{f2}}{2M^f(s)},
\end{equation}
where $E^f$ is the single-particle energy and $\Sigma_S^f$ and $\Sigma_0^f$
are the scalar and time-component self-energies~\cite{MC08}.
$V_{NR}(f)$ is fixed to be -30~MeV for $f=\Lambda$~\cite{MDG88}, 30~MeV for
$f=\Sigma$~\cite{SNA04} and -18~MeV for $f=\Xi$~\cite{Schafner-Bielich2002}. 
While the value for the $\Lambda$ is quite certain, that for the other hyperons are still under debate.
The weights $w_s^f$ are given in table~\ref{tab:weights}
while we choose $\tilde{w}_s^f=(1+s^f/3)$ for all baryons $f$.

\begin{table}[t]
\setlength{\tabcolsep}{.22in}
\renewcommand{\arraystretch}{1.5}
  \caption{Weight factor ${w}^f_s$ for the hyperons $\Lambda$, $\Sigma$ and $\Xi$.}
  \label{tab:weights}
\footnotesize\rm
  \begin{center}
  \begin{tabular}{lcccc}
    \hline\hline
    Model &$\Lambda$&$\Sigma$&$\Xi$ \\
    \hline
    MC1-H& 0.61&0.18&0.33\\
    MC1-HF&1.00&0.50&0.55 \\
    MC2-HF&1.00&0.51&0.52 \\
    MC3-HF &0.91&0.38&0.48 \\
    MC4-HF &0.89&0.39&0.47 \\
    \hline
    QMC-H&0.62&0.35&0.32 \\
    QMC-HF1&0.71&0.51&0.37 \\
    QMC-HF2&0.64&0.41&0.33 \\
    QMC-HF3&0.70&0.33&0.37 \\
\hline\hline
  \end{tabular}
  \end{center}
\end{table}

Being partly related to the underlying quark structure of baryons the chiral and QMC models are
good candidates for exploration 
of the properties of dense core of neutron stars.
In table~\ref{tab:approximations} we explore different models for the EoS with and without hyperons 
and their predictions 
for the maximum mass $M_\mathrm{max}$ and 
associated radius $R(M_\mathrm{max})$.
The first three EoS are based only on nucleons.
For the following three models in table~\ref{tab:approximations}, the EoS is based on non-interacting 
hyperons.
The maximum mass for these models is strongly reduced compared to models without
hyperons.
Interacting hyperons are taken into account in the next 10 models in table~\ref{tab:approximations}.
For the five chiral models the maximum mass is increased compared to the models with free
hyperons, but it remains smaller than that predicted by the models without hyperons.
There is a small variation of about 0.05~$M_\odot$ between the different HF models depending
on the extension of the chiral model under consideration.

\begin{table}[t]
\setlength{\tabcolsep}{.03in}
\renewcommand{\arraystretch}{1.4}
  \caption{Different models corresponding to different approximations for
nucleons (second column) and for hyperons (third column). 
The last three columns correspond to the maximum mass $M_\mathrm{max}$, its 
associated radius $R_\mathrm{max}$ and the corresponding baryonic density
$n_\mathrm{max}^c$.}
  \label{tab:approximations}
\footnotesize\rm
  \begin{center}
    \begin{tabular}{cccccc}
      \hline\hline
     EoS & Nucleon & Hyperon & $M_\mathrm{max}$ & $R(M_\mathrm{max})$ & $n_\mathrm{max}^c$\\
      &   N  & Y    & $(M_\odot) $             &  (km) & (in $n_0$) \\
      \hline
      MC1-H/N & MC1-H& No & 1.65 & 10.6 & 7.8\\ 
      MC1-HF/N & MC1-HF& No & 1.87 & 10.8 & 7.2\\ 
      MC3-HF/N & MC3-HF& No & 1.85 & 10.7 & 7.3 \\ 
      \hline
      MC1-H/NY$_{FG}$ & MC1-H& Fermi Gas& 1.02 & 11.9 & 5.6 \\ 
      MC1-HF/NY$_{FG}$ & MC1-HF& Fermi Gas& 1.12 & 12.0 & 5.4 \\
      MC3-HF/NY$_{FG}$ & MC3-HF& Fermi Gas& 1.13 & 11.9 & 5.5 \\
      \hline
      MC1-H/NY & MC1-H& MC1-H& 1.45 & 11.9 & 6.5 \\
      MC1-HF/NY & MC1-HF& MC1-HF& 1.58 & 12.2 & 5.6 \\
      MC2-HF/NY & MC2-HF& MC2-HF& 1.60 & 12.2 & 5.5 \\
      MC3-HF/NY & MC3-HF& MC3-HF& 1.55 & 12.0 & 5.6 \\
      MC4-HF/NY & MC4-HF& MC4-HF& 1.55 & 12.0 & 5.6 \\
      \hline
      QMC700/NY & QMC700 & QMC700 & 2.00 & 12.5 & 5.1 \\
      QMC-H/NY & QMC-H & QMC-H & 1.68 & 12.2 & 5.5 \\
      QMC-HF1/NY & QMC-HF1 & QMC-HF1 & 1.97 & 12.3 & 5.2\\
      QMC-HF2/NY & QMC-HF2 & QMC-HF2 & 1.97 & 13.3 & 4.6\\
      QMC-HF3/NY & QMC-HF3 & QMC-HF3 & 1.66 & 12.3 & 5.3\\
      \hline\hline
    \end{tabular}
  \end{center}
\end{table}

In the last four lines of table~\ref{tab:approximations} we present variants of the QMC model whose
parameters are given in table~\ref{tab:parameters}. 
The original QMC700/NY model predicts a maximum mass of 2.00~M$_\odot$ with hyperons~\cite{RGM08}.
The model QMC-H/NY is a consistent Hartree model where the $s$ and $\omega$ coupling constants
are readjusted to reproduce saturation properties ($n_0$, $B_0$ and $V_{NR}(f)$).
The maximum mass in the model QMC-H/NY is lower than that of the original QMC700/NY since the
incompressibility is lower.
In the model QMC-HF1/NY the Hartree and Fock terms are consistently treated in the mean-field potential
and the same readjustment procedure as in QMC-H/NY is followed. 
The Fock contribution to the total energy is however treated non-relativistically  as in the QMC700/NY model.
Including the complete Fock contribution in the energy, we obtain the fully relativistic QMC-HF2/NY model.
The maximum mass is not altered by the fully relativistic 
treatment of the Fock term. 
Finally, we reduced the incompressibility by increasing the parameter $C$ from $0.51$ to $1$
(see table~\ref{tab:parameters}) and the maximum mass decreases from 1.97 to 1.66~M$_\odot$.
From this analysis, we conclude that the large mass obtained in the QMC700/NY model~\cite{RGM08} 
is mainly due to an overestimated and incompressibility.

\begin{figure}[t]
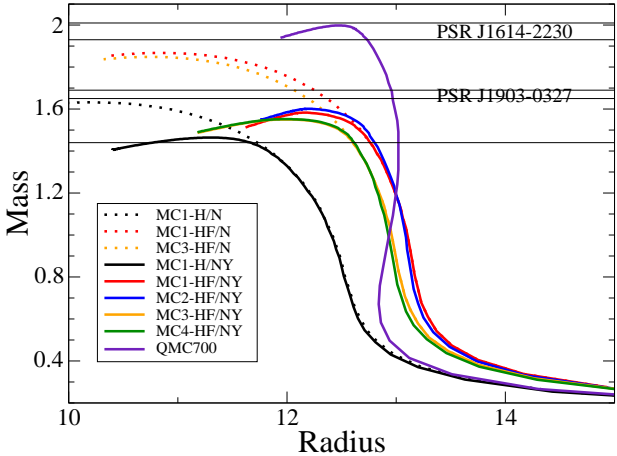

\onefigure[scale=0.35]{RM.eps}
\caption{Mass of neutron stars versus the radius for selected EoS.}
\label{fig:mr}
\end{figure}

We show in figure~\ref{fig:mr} the mass-radius relation for the various EoS
presented in this Letter.
Dotted lines show results for the models without hyperons while the solid lines stand for the models
including interacting hyperons.
As expected the EoS with hyperons reach lower maximum masses than the models without hyperons,
except the model QMC700 due to its large and unrealistic incompressibility.
The maximum mass obtained with hypernuclear EoS does not change much for different models. 
This is due to a self-regulating compensation effect between the softening of the EoS and the onset of the 
hyperonic degree of freedom~\cite{Schulze2006,Schulze2011}.
The maximum mass of about 1.6~$M_\odot$ predicted by the relativistic approach is slightly larger than
that of 1.4~$M_\odot$ found in Ref.~\cite{Schulze2011}.
The difference of 0.2~$M_\odot$ is mainly related to the larger incompressibility of the relativistic models.
It is also observed that the radii for the Hartree models are smaller than that of the Hartree-Fock ones,
and that the predicted radii of the relativistic models and that of Ref.~\cite{Schulze2011} are comparable.

\begin{figure}[t]
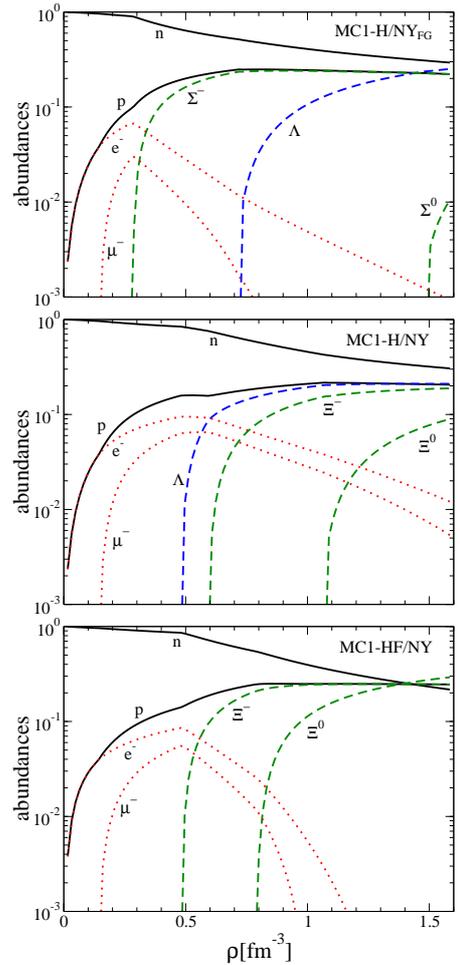

\onefigure[scale=0.25]{abondances_model1.eps}
\onefigure[scale=0.25]{abondances_model4.eps}
\onefigure[scale=0.25]{abondances_model7.eps}
\caption{Particules abundances for the various models of table~\ref{tab:approximations}:
MC1-H/NY$_{FG}$ (top), MC1-H/NY (middle) and MC1-HF/NY (bottom). }
\label{fig:abundances}
\end{figure}

The abundances for nucleons and hyperons are shown for various EoS in 
figure~\ref{fig:abundances}: with free hyperons (MC1-H/NY$_{FG}$) 
and with interacting hyperons (MC1-H/NY and MC1-HF/NY).
The order of appearance of hyperons is different for these models and depends strongly on the
corresponding interactions. 
In the model MC1-H/NY$_F$ with free hyperons, the  $\Sigma^-$ is non interacting
and appears first. On the other hand, in the models MC1-H/NY and MC1-HF/NY,
the $\Sigma^-$ interaction is repulsive while the $\Xi^-$ interaction is attractive,
therefore the latter tends to prevent the $\Sigma^-$ from appearing. 
In the model MC1-H/NY, the $\Lambda$ appears first followed closely in density
by the $\Xi^-$, then at a larger density by the $\Xi^0$, while for the model MC1-HF/NY 
the $\Xi^-$ and $\Xi^0$ appear first and push the other hyperons to larger densities.

We have presented in this Letter different EoS for hyperon matter derived 
from both chiral and QMC models.
These models are thermodynamically consistent and treat the Hartree and Fock terms on equal footing.
We have found that the different chiral parameterizations give a maximum neutron
star mass that does not exceed 1.6~M$_\odot$ for HF models with $K_0\approx 270$~MeV.
Correcting the QMC700 model~\cite{RGM08} by including the complete relativistic
contribution to the Fock term, a correct incompressibility and the experimental
constraints for the hyperon mean-fields, we found a decrease of the predicted maximum mass from 
2.00 down to 1.66~M$_\odot$. 
We additionally checked that the effect of rotation with the period of 3.15~ms (as for PSR J1614-2230)
does not increase the maximum mass by more than 0.02~$M_\odot$.
We therefore conclude that it is difficult to reconcile any of the model presented in this work 
with the observed mass of PSR J1614-2230 and the empirical knowledge of saturation properties of
nuclear matter. 
Unless an unexpected property of the hyperon interaction is missing in the present model, as well as in
the BHF models of Refs.~\cite{Baldo2000,Schulze2006,Schulze2011,vidana2011}, our results tends
to exclude hypernuclear matter to be present in the core of massive neutron stars.
It reinforces alternative models such as for instance the model of deconfined quark matter, for which
experimental constraints are almost inexistent.
We can therefore conclude that the equation of state of dense matter in the core of neutron star is 
still not even qualitatively understood.

\acknowledgments
We thank I. Vida\~na for fruitful discussions during the completion of this work.
This work was partially supported by the ANR NExEN and SN2NS project ANR-10-BLAN-0503 
contracts and COMPSTAR, an ESF Research Networking Program.



\end{document}